\documentstyle[bezier,emlines2,12pt]{article}

\title{\bf Inclusive jet production on the nucleus in the
perturbative QCD with $N_c\rightarrow\infty$}
\author{M.Braun \\ Department of High Energy physics,
 University of S.Petersburg,\\
198904 S.Petersburg, Russia}
\def\beq{\begin{equation}}
\def\eeq{\end{equation}}
\def\noi{\noindent}
\begin{document}
\maketitle
\medskip
\noi{\bf Abstract.}
Using the recently found hA amplitude,
single and double inclusive jet production rates on the nucleus
are studied in the perturbative QCD with $N_c\rightarrow\infty$.
Jet multiplicities are
found to grow with $A$ and energy as $A^{2/9}s^\Delta$ where $\Delta$
is the BFKL intercept. Long range correlations are found to
behave as $A^{4/9}s^{2\Delta}$ and change from negative to positive
at a certain high energy. 
\vspace{3.5cm}

\noi{\Large\bf SPbU-IP-00-07}

\newpage
\section{Introduction}
In the  colour dipole approach of A.H.Mueller [1,2] with
$N_c\rightarrow\infty$ the interaction 
with a heavy nucleus is exactly described by the sum of fan diagrams
constructed 
of BFKL pomerons, each of them splitting into two [3,4]. 
The equation for this sum [3-6] has recently been studied perturbatively
in [7], by asymptotic estmates in [8] and finally solved numerically in [6].
The results indicate that the cross-section of a particle on the nucleus
saturates at high energies to its geometrical value $2\pi R_A^2$
corresponding to the scattering on a black disk. In itself this is not a
very surprising result: since the days of the old Regge-Gribov theory it
has been known that the supercritical pomeron fan diagrams with a triple
pomeron vertex lead
to a constant cross-section [9]. The new elements in the recent
developments
is  that it has been understood that in the high-colour limit no splitting
of the pomeron in
more than two occurs and that the fan diagrams seem to be unitary by
themselves so that they indeed represent the full scattering amplitude on
the nucleus.

An interesting byproduct of the numerical solution in [6] is the found
gluon density of the nucleus in the combined momentum-impact parameter
space 
\beq
\frac{\partial xG(x,q,b)}{\partial^2q\partial^2b}=
\frac{N_c}{2\alpha_s\pi^2}h(y,q,b)
\eeq
where the rapidity $y$ is related to $x$ by
\[ y=\ln\frac{1}{x}
\]
Function $h$ defined via the sum of BFKL fan diagrams below (Section 2)
and found in [6] numerically
is a soliton wave in $y-\ln q$ space moving
towards higher rapidities with a constant velocity and preserving its
nearly Gaussian shape. A fit to numerical data gives
\beq
h(y,q,b)=h_0e^{-a(\xi-\xi_0(y,b))^2}
\eeq
where $\xi=\ln q$, $h_0\simeq 0.3$ and $a\simeq 0.3$ are universal and
$\xi_0$ is nearly linear in $y$:
\beq
\xi_0=c(b)+\Delta_0 y,\ \ \Delta_0=2.3\bar{\alpha}
\eeq
where we standardly define $\bar{\alpha}=\alpha_sN_c/\pi$.
The term $c(b)$ actually depends on a dimensionless parameter $B$:
\[ B=\pi\alpha_s^2AT(b)R_N^2\]
where $R_N$ is the nucleon radius and $T(b)$ is the nucleus profile
function. Numerical results for $c(b)$ are  well described by
\[ c(b)=-3.11+(2/3)\ln B\]
Actually the gluon density does not really show itself in the total
scattering cross-section at high rapidities. As mentioned, the
cross-section becomes purely geometrical and insenstive to any dynamics.
One expects however that the gluon density will be directly felt in
production processes, where it is responsible for hard collisions giving
rise to observable jets.

In this paper we study  simplest of the production processes: the
single and double inclusive jet production in hA collisions.
It has to be noted that
these observables were also studied for the fan diagrams
in the old Regge-Gribov theory.
The most important conclusion which was drawn from this study is that
the multiplicities are strongly damped, as compared to the naive eikonal
result. For hA scattering one obtains a multiplicity independent of $A$
(instead of $\sim A^{1/3}$).
As we shall see this result does not hold
for the BFKL fan diagrams: the jet production rate grows
with $A$ although somewhat slowlier than in the eikonal approach,
approximately as $A^{2/9}$.
The energy dependence remains esentially the same as in older
studies: the production rate grows with rapidity $Y$
essentially as $\exp \Delta Y$, where $\Delta$ is the BFKL intercept.
With $\alpha_s\sim 0.2$ its value is close to 0.5, so that the
multiplicities reach very high values at rapidities in the region
$20\div 30$.
\section{Single jet inclusive production}
In the BFKL framework, at fixed impact parameter $b$, 
a single scattering contribution to the forward ampitude
for the interaction of the projectile particle with a nucleus
has a form
\beq{\cal A}_1(Y,b)=isg^4AT(b)\int d^2r_1d^2r_2\rho(r)G(Y,r_1,r_2)
\rho_N(r_2)\equiv 2is\int d^2r_1\rho(r_1)\Phi_1(y,r_1,b)\eeq
Here $\rho$ and $\rho_N$ are the colour densities of the projectile
and the target nucleon respectively. Function $G$ is the forward
BFKL Green function [10] taken at an overall rapidity $Y$
\beq
G(Y,r,r')=\frac{rr'}{32\pi^2}
\sum_{n=-\infty}^{+\infty}e^{in(\phi-\phi')}
\int_{-\infty}^{\infty}\frac{d\nu e^{Y\omega(\nu)}}
{[\nu^{2}+(n-1)^{2}/4][\nu^2+(n+1)^2/4]}(r/r')^{-2i\nu},
\eeq
where $\phi$ and $\phi'$ are the azimuthal angles and
\beq
\omega(\nu)=\bar{\alpha}(\psi(1)-{\rm Re}\psi(1/2+i\nu))
\eeq
are the BFKL levels.
Due to  the azimuthal symmetry of the projectile colour
density one may retain only the term with zero orbital momenta
$n=0$ in (5).
The corresponding inclusive jet production rate
\beq
I(y,k,b)\equiv\frac{(2\pi)^2\partial\sigma}{\partial y\partial^2k\partial ^2b}
\eeq
is obtained from the imaginary part of (4) divided by $s$ after the
substitution
\beq
G(Y,r_1,r_2)\rightarrow \int d^2rG(y_1,r_1,r)V_k(r)G(y,r,r_2)
\eeq
where $r$'s are relative distances between the gluons and
$V_k(r)$ is a vertex for the emission
\beq
V_k(r)=\frac{4N_c\alpha_s}{k^2}\stackrel{\leftarrow}{\Delta} e^{ikr}
\stackrel{\rightarrow}{\Delta}
\eeq
local in $r$ (a differential operator, the arrows shows the direction of its
action). The rapidity of the produced jet relative to the projectile
is  $y_1=Y- y$.

As mentioned, in the limit $N_c\rightarrow\infty$ the total
scattering amplitude which includes multiple interactions in the target is 
given by the sum of fan diagrams 
shown in Fig 1. To obtain the corresponding inclusive cross-section, 
one has to make the substitition (8) in one of the pomerons. However due
to
the well-known AGK rules all such substitutions in the pomerons
below the upper splitting point (Fig. 2a) cancel, since the other branch,
counting from the upper point, can be both "cut" and "uncut" (i.e.
represent both real and virtual processes). The only left contribution is
the one with the substitution (8) in the uppermost pomeron (Fig. 2b).
It is clearly seen from Fig. 2b  that the vertex $V_k$ is coupled
from below to exactly the full amplitude $\Phi(y,r,b)$ which
multiplies the projectile density in Fig. 1 and substitutes the single
pomeron exchange contribution $\Phi_1(Y,r_1,b)$ in (4). 
Thus one obtains the
inclusive cross-section for jet production on the nucleus at fixed $b$
as
\beq
I(y,k,b)=2\int d^2r'd^2r\rho(r')G(y_1,r',r)V_k(r)\Phi(y,r,b) 
\eeq
The total inclusive production rate as a function of $y$ and $k$ is
obtained from (10) after the integration over $b$

Function $\phi(y,r,b)= \Phi(y,r,b)/(2\pi r^2)$, in the momentum space,
satisfies
a nonlinear equation [6]
\beq
\frac{\partial\phi(y,q,b)}{\partial
\tilde{y}}=-H\phi(y,q,b)-\phi^2(y,q,b),
\eeq
where  $\tilde{y}=\bar{\alpha}y$ and $H$ is the BFKL Hamiltonian for the
so-called semi-amputated function
[10]. As mentioned in the Introduction, this equation was solved numerically
in [6].

The Laplace operator in (9) acting on $\Phi(y,r,b)$ gives
\[
\nabla_r^2\Phi(y,r,b)=2\pi\nabla_r^2 r^2\int \frac{d^2q}{(2\pi)^2}e^{iqr}
\phi(y,q,b)=-2\pi\nabla_r^2
\int \frac{d^2q}{(2\pi)^2}e^{iqr}\nabla_q^2\phi(y,q,b)
\]\beq=2\pi
\int \frac{d^2q}{(2\pi)^2}e^{iqr}q^2\nabla_q^2\phi(y,q,b)=
2\pi h(y,r,b)
\eeq
Note that in the integration by parts used to obtain the second
expression one has to be careful, since function $\phi(y,q,b)$ has a
logarithmic singularity at $q=0$ of the form $-\ln q$. As a result, action
of the Laplacian operator in $q$ generates a $\delta^2(q)$ term which in
its turn leads to a constant in $\Phi$ (actually unity). However this term
is annihilated by the Laplacian operator in $r$, so that no extra term
arises from the intergration by parts.

Function $h(y,q,b)$, which is the gluon density, up to a trivial factor 
(Eq. (1)) is just a Fourier transform of $h(y,r,b)$:
\beq h(y,q,b)=q^2\nabla_q^2\phi(y,q,b)\eeq
So we find the inclusive cross-section in the form
\beq
I(y,k,b)=\frac{16\pi\alpha_sN_c}{k^2}\int
d^2r'd^2r\rho(r')
\nabla^2_rG(y_1,r',r)e^{irk}h(y,r,b)
\eeq
We observe that the definition of $h(y,q,b)$ as the gluon density
introduced in [6] is supported by its role in the jet production on the
nucleus.

Applying the Laplace operator to the Green function (5) (with only $n=0$
terms retained) we obtain
\beq
\nabla_r^2G(y_1,r',r)=-\frac{r'}{8\pi^2 r}
\int_{-\infty}^{\infty}\frac{d\nu e^{y_1\omega(\nu)}}
{(\nu+i/2)^2}(r'/r)^{-2i\nu},
\eeq
We are going to study the production rate at $y_1>>1$, that is far from
the projectile fragmentation region. Then we can use the asymptotics of
(15)
at high $y_1$, which comes from $\nu\sim 0$. Presenting $\omega (\nu)$ in
the standard manner
\beq
\omega(\nu)=\Delta y-\beta\nu^2
\eeq
where
\beq
\Delta=\bar{\alpha}\,4\ln 2,\ \ {\rm and}\ \ 
\beta=\bar{\alpha}\,14\zeta(3)
\eeq
we get for (15)
\beq
\nabla_r^2G(y_1,r',r)\simeq\frac{r'}{2\pi^2 r}e^{\Delta
y_1}\sqrt{\frac{\pi}{\beta y_1}}
\eeq

With (18), integration over $r'$ converts $r'$ into
the average transverse dimension of the projectile $\langle r\rangle_P=R_P$.
Doing the integration over $r$ we obtain
\beq
I(y,k,b)=\frac{8\bar{\alpha}}{k^2}
R_Pe^{\Delta y_1}\sqrt{\frac{\pi}{\beta
y_1}}F(y,k,b)
\eeq
where
\beq
F(y,k,b)=\int\frac{d^2r}{r}e^{ikr}h(y,r,b)=
\int\frac{d^2q}{2\pi}\frac{h(y,q,b}{|k+q|}
\eeq

As mentioned in the Introduction, the gluon density $h(y,q,b)$ has a
sharp peak at
$q=q_0(y,b)=\exp\xi_0(y,b)$ and is practically zero everywhere else.
With growing $y$ $q_0(y,b)$ grows exponentially, so that at high $y$
we can safely neglect $k$ in the denominator of (20) as compared
to $q$. Then $F$ becomes independent of $k$ and given by
\beq
F(y,b)=\int\frac{d^2q}{2\pi q}
h(y,q,b)=
\int_{-\infty}^{+\infty} d\xi\, q\, h(y,q,b)
\eeq
where  $\xi=\ln q$.
One easily finds that
\beq
\int d\xi h(y,q,b)=1
\eeq
Indeed Eq. (13) can be rewritten as
\beq
h(y,q,b)=\frac{d^2}{d\xi^2}\phi(y,q,b)
\eeq
Putting (23) into (21) and taking into account that at $\xi\rightarrow
+\infty$
$\phi=const-\xi$ and at $\xi\rightarrow +\infty$ $\phi\rightarrow 0$
we obtain (22). So (21) in fact has a meaning of the average gluon
momentum in the nucleus $\langle q\rangle$. Since
$h$ has a sharp maximun at $\xi=\xi_0(y,b)$ given by (3) one expects that
$F(y,b)$
will approximately be given just by $q_0(y,b)$. Numerical integration gives
more accurate values which are well fitted by the formula similar to (3)
but with  slightly different parameters:
\beq
F(y,b)=q_1(y,b)=1.2 B^{2/3}\frac{e^{\Delta_1y}}{R_N\sqrt{y}},\
\  \ \Delta_1=2.38\bar{\alpha}
\eeq

With this approximate form of $F$ we find
the final expression for the inclusive rate production (in (GeV/c)$^{-2}$) 
\beq
I(y,k,b)=1.2\frac{8\bar{\alpha}}{k^2}\frac{R_P}{R_N}
[\pi\alpha_s^2AT(b)R_N^2]^{2/3}e^{\Delta Y-\epsilon y}
\sqrt{\frac{\pi}{\beta y(Y-y)}}
\eeq
where
\[\epsilon=\Delta-\Delta_1=0.39\bar{\alpha}\]
and is relatively small. As we observe, the
inclusive
jet production is maximal in nucleus fragmentation region and slowly falls
as one moves to higher $y$.

Since the total hA cross-section saturates at large $Y$ an immediate
consequence of (25) is that the multiplicity grows like $\exp\Delta Y$,
that
is, as the cross-section generated by a single pomeron exchange.  This
conclusion is also true for the inclusive production from the sum of fan
diagrams in the  old Regge-Gribov theory with a local pomeron. However
this theory predicts a much stronger $y$-dependence, corresponding to
(25) with $\epsilon=\Delta$.

As to the $A$-dependence, a novel element in the BFKL fan diagrams is 
factor $[\pi\alpha_s^2 AT(b)R_N^2]^{2/3}$.
For a nucleus with
constant density inside a sphere of radius $R_A=A^{1/3}R_0$ 
its integration over all $b$ leads to a factor
\beq
2\pi R_A^2\,A^{2/9}\frac{3}{8}\gamma,\ \ \gamma=
\left(\frac{3}{2}\frac{R_N^2}{R_0^2}\right)^{2/3}
\eeq
Since the total hA cross-section at large $Y$ saturates to its black-disc
limit $2\pi R_A^2$ the jet multiplicity results
proportional to $A^{2/9}$:
\beq
\mu(y,k)=\frac{1}{\sigma^{tot}_{hA}}\int d^2bI(y,k,b)=
1.2A^{2/9}\frac{3\bar{\alpha}}{k^{2}}\frac{R_P}{R_N}\gamma
e^{\Delta Y-\epsilon y}
\sqrt{\frac{\pi}{\beta y(Y-y)}}
\eeq
So unlike the fan diagrams in the old Regge-Gribov theory,
the BFKL model predicts rising of
the multiplicities with $A$, although somewhat weaker than according to
the eikonal model ($\propto A^{1/3}$).

Although the gluon distribution in the nucleus results radically changed as
compared to the single nucleon, in particular, strongly damped at
small momenta, this  seems to have no effect on the $k$-dependence
of the jet production rate at moderate $k<<q_1(y,b)$: it remains
proportional to $1/k^2$ and  growing as $k\rightarrow 0$.
The inclusive cross-section integrated over all $k$ remains
infinite, as for a single BFKL pomeron exchange. Thus discussing the
integrated cross-section we have to cutoff the spectrum from below by some
$k_{min}$. The formal expression for this cross-section can be trivially
found integrating (19) over all $k>k_{min}$. A simple estimate of its
magnitude can be made in the logarithmic approximation in the large
parameter $q_1(y,b)/k_{min}$. Evidently
\[
\int\frac{d^2k}{(2\pi)^2k^2}F(y,k,b)\theta(k-k_{min})=
\int\frac{d^2q}{2\pi}h(y,q,b)\int\frac{d^2k}{(2\pi)^2k^2}\frac{1}{|k+q|}
\theta(k-k_{min})\]\beq
\simeq\frac{1}{2\pi}\int\frac{d^2q}{2\pi q}h(y,q,b) \ln(q/k_{min})=
\frac{1}{2\pi}\langle q\ln(q/k_{min})\rangle
\eeq
So, up to factor $1/2\pi$, integration over $k$ leads to the average value
of $q\ln(q/k_{min})$ in the nucleus. Approximating it 
by $q_1(y,b)\ln (q_1(y,b)/k_{min})$, so that for fixed $b$ one finds
\beq
I(y,b)\equiv\int\frac{d^2k}{(2\pi)^2}\theta(k-k_{min})=
\frac{4\bar{\alpha}}{\pi}
R_Pe^{\Delta (Y-y)}\sqrt{\frac{\pi}{\beta (Y-y)}}
q_1(y,b)\ln\frac{q_1(y,b)}{k_{min}}
\eeq
with $q_1(y,b)$ given by (24).

At high $y$ the logarithmic factor reduces to $\Delta_1 y$.
Its effect, opposite to the exponential, is to flatten
the distribution $I(y)$ as
compared to $I(y,k)$. The $A$ dependence evidently becomes somewhat stronger.
Numerical integration of $I(k,y,b)$ with function $h$ found in [6]
gives results which more or less agree with this approximate estimates.
At $\bar{\alpha}Y=6$ for lead ($A=207$) we obtain multiplicities
which in the region $2\leq \bar{\alpha}y\leq 5$ very slowly grow from
$1.13\,10^5$ to $1.26\,10^5$. The found $A$ dependence can be fitted by
$\sim A^{0.26}$. The large magnitude of the multiplicity is due to
the exponential factor $\exp \Delta Y=\exp (\bar{\alpha}Y 4\ln 2)$.

\section{Double inclusive cross-sections}
Passing to double inclusive cross-sections and using the AGK rules we
find two contributions schematically shown in Figs. 3 a,b.

A simpler contribution Fig. 3a corresponds to emission of both gluons
from the upper pomeron. To obtain it we have to make two insertions
(8) in the upper BFKL Green function. We get
\[
I_a(y_1,k_1,y_2,k_2,b)\equiv\Big[\frac{(2\pi)^4\partial\sigma}
{\partial y_1\partial^2k_1\partial y_2\partial d^2k_2\partial ^2b}
\Big]_{Fig. 3a}\]\beq=
2\int d^2r'd^2r_1d^2r_2\rho(r')G(Y-y_1,r',r_1)V_{k_1}(r_1)
G(y_1-y_2,r_1,r_2)V_{k_2}(r_2)\Phi(y_2,r_2,b)
\eeq
where we assumed that $Y>>y_1>>y_2>>1$ in the lab. system.

The only new
element in this equation is the second BFKL Green function with
Laplacians applied in both coordinates. Using (5) we find
\beq
\Delta_1\Delta_2G(y,r_1,r_2)=\frac{1}{2\pi^2r_1r_2}
\int d\nu e^{y\omega(\nu)}\left(\frac{r_1}{r_2}\right)^{-2i\nu}
\eeq
Combining this with (15) for the first Green function
(with the integration variable $\nu_1$) we find an integral over $r_1$:
\beq
\int d^2r_1 e^{ik_1r_1}r_1^{-2-2i\nu+2i\nu_1}=
\pi (k_1/2)^{2i(\nu-\nu_1)}\frac{1}{i(\nu_1-\nu-i0)}
\frac{\Gamma(1+i\nu_1-i\nu)}{\Gamma(1-i\nu_1+i\nu)}
\eeq
As before we limit ourserlves with the leading contribution as
both $Y-y_1$ and $y_1-y_2$ become large, which correspond to small
values of both $\nu$ and $\nu_1$. Correspondingly we put $\nu=\nu_1=0$
everywhere except in the exponents $\omega(\nu_1)(Y-y_1)$ and
$\omega(\nu)(y_1-y_2)$ and in the denominator of (32). We find that only
the real part of (32) contributes, which is equivalent to taking
\[\frac{1}{i(\nu_1-\nu-i0)}\rightarrow \pi\delta(\nu_1-\nu)\]
Using (16) we finally get for the double inclusive cross-section (30)
\beq
I_a(y_1,k_1,y_2,k_2,b)=16\pi R_P\frac{\bar{\alpha}^2}{k_1^2k_2^2}
e^{\Delta(Y-y_2)}\sqrt\frac{\pi}
{\beta(Y-y_2)}F(y_2,k_2,b)
\eeq
where function $F(y,k,b)$ was defined in Section 2 (Eq.(20)).
A remarkable feature of this cross-section is that it depends only on
the rapidity of the slowlier jet $y_2$ and does not
depend on $y_1$ altogether.  In the approximation (24) the dependence
on $y_2$ results the same as for the single inclusive cross-section, namely
$\exp (-\epsilon y_2)$, so that the cross-section slowly diminishes as
$y_2$ increases.

Using approximation (24), integrating over $b$ and dividing by
$\sigma_{hA}^{tot}$ we get the corresponding multiplicity distribution
\beq
\mu_a(y_1,k_1,y_2,k_2)=1.2 A^{2/9}6\pi\frac{R_P}{R_N}\gamma
\frac{\bar{\alpha}^2}{k_1^2k_2^2}
e^{\Delta Y-\epsilon y_2}\sqrt\frac{\pi}
{\beta y_2(Y-y_2)}
\eeq
where $\gamma$ was defined in (26).

The second contribution (Fig. 3b) corresponds to emission of both jets
just after the first splitting of the upper pomeron. To write it we
first
take the contribution from all non-trivial fan diagrams to $\Phi$ [6]
\[
\Phi(Y,r',b)=
-\frac{g^2N_c}{8\pi^3}
\int_0^Y dy' \int\prod_{i=1}^3 d^2x_i
\delta^2(x_1+x_2+x_3)\]\beq\frac{x_3^2{\nabla_3}^4}{x_1^2x_2^2}
G(Y-y',r',x_3)\Phi(y',x_1,b)\Phi(y',x_2,b).
\eeq
where we denoted $x_i,\ i=1,2,3$ the transverse coordinates of the
triple pomeron vertex.
According to the AGK rules the corresponding contribution to the
absorptive part with both lower branches cut is twice (35) with the opposite
sign. To finally obtain the double inclusive cross-section we have
to insert into both $\Phi$'s on the right-hand side verteces $V_k(r)$
as was done in (10). In this way we obtain the second part of the double
inclusive cross-section as
\[
I_b(y_1,k_1,y_2,k_2,b)=\frac{g^2N_c}{2\pi^3}
\int d^2r'\rho(r')d^2r_1d^2r_2
\int_{y_1}^Y dy' \int\prod_{i=1}^3 d^2x_i
\delta^2(x_1+x_2+x_3)\]\beq\frac{x_3^2{\nabla_3}^4}{x_1^2x_2^2}
G(Y-y',r',x_3)G(y'-y_1,x_1,r_1)V_{k_1}(r_1)\Phi(y_1,r_1,b)
G(y-y_2,x_2,r_2)V_{k_2}\Phi(y_2,r_2,b).
\eeq
We again assume $Y>>y_1>>y_2>>1$ so that the
rapidity  of the splitting point $y'$ varies from $y_1$ to $Y$.

We easily find
\beq
x_3^2\nabla_3^4G(y,r',x_3)=\frac{r'}{2\pi^2 x_3}
\int d\nu_3e^{y\omega(\nu_3)}\left(\frac{r'}{x_3}\right)^{2i\nu_3}
\eeq
Using this and representation (15) for the other two BFKL Green functions
we find an integral over $x_i$, $i=1,2,3$ 
\beq
J(\nu_i)=
 \int\prod_{i=1}^3 d^2x_i\delta^2(x_1+x_2+x_3)
 \frac{1}{x_1x_2x_3}x_1^{-2i\nu_1}x_2^{-2i\nu_2}x_3^{-2i\nu_3}
\eeq
This integral diverges at large $x_i$'s for real $\nu$'s.
In fact, presenting the $\delta$ function as an integral over  momentum $q$
we find
\beq
J=2\pi\prod_{l=1}^3 2^{-2i\nu_l}\frac{\Gamma(1/2-i\nu_l)}
{\Gamma(1/2+i\nu_l)}\int\frac{d^2q}{q^3}q^{2i(\nu_1+\nu_2+\nu_3)}
\eeq
which does not exist for real $\nu$ due to the divergence at $q=0$.
To overcome this difficulty we have to shift the integration  contours
in $\nu$'s into the lower half-plane, using the fact that integrands are
analytic in the strip $-i/2<{\rm Im}\,\nu<+i/2$. We choose to do it
in a symmetric manner putting
\beq
\nu_i=-i/6+\bar{\nu}_i,\ \ i=1,2,3
\eeq
With such a choice we get an extra factor $q$ in the integral over $q$,
which then gives a $\delta$ function:
\beq
J=\pi^2\prod_{l=1}^3 \frac{\Gamma(1/2-i\nu_l)}
{\Gamma(1/2+i\nu_l)}\delta(\bar{\nu}_1+\bar{\nu}_2+\bar{\nu}_3)
\eeq

Now we pass to the integration over the rapidity of the splitting point.
The integral is done trivially:
\beq
\int_{y_1}^{Y}dy'e^{y(\omega_1+\omega_2-\omega_3)}=
\frac{e^{Y(\omega_1+\omega_2-\omega_3)}-e^{y_1(\omega_1+\omega_2-\omega_3)}}
{\omega_1+\omega_2-\omega_3}
\eeq
where we denote $\omega_1=\omega(\nu_1)$ etc. Multiplying by the rest
exponential factors we find the overall rapidity factor as
\beq
\frac{e^{(Y-y_1)\omega_1+(Y-y_2)\omega_2}
-e^{(Y-y_1)\omega_3+(y_1-y_2)\omega_2}}
{\omega_1+\omega_2-\omega_3}
\eeq
Note that each of the two terms depends only on two of the three $\nu$
variables. Therefore integrating over $\nu$'s we can always take as
independent variables just those which are present in the exponents
in (43).

Having this in mind we again use the fact that all rapidity differences
in the exponents are large and use the saddle point method. Athough
the initial contours of integration in the independent $\nu$'s lie
along the lines ${\rm Im}\,\nu=-1/6$, the saddle points  remain at
$\nu=0$, that is at $\bar{\nu}=(1/6)i$. This means that independent
$\nu$'s will be close to zero, as usual. The only price we have to pay for
the shift in the contours of integration will be the value of the
dependent $\nu$ at the saddle point.

Take the first term in (43), for which
$\nu_1$ and $\nu_2$ are independent variables and $\nu_3$ is the
dependent one. At the saddle point we shall have $\nu_1=\nu_2=0$
and $\nu_3=-(1/6)i-\bar{\nu_1}-\bar{\nu_2}=-(1/2)i$. 
Accordingly we put $\nu_1=\nu_2=0$ and $\nu_3=-(1/2)i$ in all terms
except in the exponent and the singular $\Gamma$-function:
\beq
\Gamma(1/2-i\nu_3)=\Gamma(i\nu_1+i\nu_2)=\frac{\Gamma(1+i\nu_1+i\nu_2)}
{i(\nu_1+\nu_2-i0)}\simeq\frac{1}{i(\nu_1+\nu_2-i0)}
\eeq
Again only the real part contributes, so that (43) actually gives
\[ \pi\delta(\nu_1+\nu_2)\]
Using (16) and integrating over $\nu_1$ and $\nu_2$
we  obtain a factor
\beq
\frac{1}{\Delta} e^{\Delta(2Y-y_1-y_2)}
\sqrt{\frac{\pi}{\beta(2Y-y_1-y_2)}}
\eeq
and an extra factor $r'$ due to the fact that $\nu_3=-(1/2)i$ at the saddle
point.

For the second term in (43) the independent variables are
$\nu_2$ and $\nu_3$
which are zero at the saddle point. The dependent $\nu_1$ is equal to
$-(1/2)i$ at the saddle point. The $\Gamma$  functions in (41) will
 give  $\pi\delta(\nu_2+\nu_3)$, so that the integration over $\nu_2$
 and $\nu_3$ will give a factor
\beq
\frac{1}{\Delta} e^{\Delta(Y-y_2)}\sqrt{\frac{\pi}{\beta(Y-y_2)}}
\eeq
and we shall have an extra factor $r_1$ due to $\nu_1=-(1/2)i$ at the
saddle point.

Other factors in (36) are straightforward.
Combining all of them  we find the expression
for the double inclusive cross-section corresponding to Fig. 2b in the
form
\[
I_b(y_1,k_1,y_2,k_2,b)=
\frac{4}{\ln 2}\frac{\bar{\alpha}^2}{k_1^2k_2^2}
e^{\Delta(2Y-y_1-y_2)}\sqrt{\frac{\pi}{\beta(2Y-y_1-y_2)}}
F(y_2,k_2,r)\]\beq
\Big[\langle r^2\rangle_P F(y_1,k_1,b)-
R_Pe^{-\Delta(Y-y_1)}\sqrt{\frac{2Y-y_1-y_2}{Y-y_2}}\,h(y_1,k_1,b)\Big]
\eeq
where
\beq
\langle r^2\rangle_P=\int d^2r r^2\rho(r)
\eeq
The second term is evidently exponentially small as compared to the first
and we can safely neglect it, having in mind all approximation already made.
So our final expression is
\[
I_b(y_1,k_1,y_2,k_2,b)=\]\beq
\frac{4}{\ln 2}\langle r^2\rangle_P
\frac{\bar{\alpha}^2}{k_1^2k_2^2}
e^{\Delta(2Y-y_1-y_2)}
\sqrt{\frac{\pi}{\beta(2Y-y_1-y_2)}}
 F(y_1,k_1,b)F(y_2,k_2,r)
\eeq
Recalling our expression (19) for the single inclusive cross-section we
can rewrite it as
\beq
I_b(y_1,k_1,y_2,k_2,b)=
\frac{1}{16\ln 2}\frac{\langle r^2\rangle_P}
{\langle r\rangle_P^2}
\sqrt{\frac{\beta}{\pi}\frac{(Y-y_1)(Y-y_2)}{2Y-y_1-y_2}}
I(y_1,k_1,b)I(y_2,k_2,b)
\eeq

After the integration over $b$ with the approximation (24)
we get the corresponding multiplicity distribution:
\beq
\mu_b(y_1,k_1, y_2, k_2)=1.44 A^{4/9}\frac{6}{5\ln 2}
\frac{\langle r^2\rangle_P}{R_N^2}\gamma^2
\frac{\bar{\alpha}^2}{k_1^2k_2^2}
e^{2\Delta Y-\epsilon(y_1-y_2)}
\sqrt{\frac{\pi}{\beta y_1y_2(2Y-y_1-y_2)}}
\eeq
Evidently at high $Y$ the part $\mu_b$ dominates due to the
factor $\exp 2\Delta Y$ as compared with only $\exp \Delta Y$ in
$\mu_a$

Correlations are determined by the ratio
\beq
R(Y,y_1,y_2)=\frac{\mu(y_1,k_1,y_2,k_2)}{\mu(y_1,k_1)\mu(y_2,k_2)}
\eeq
(evidently it does not depend on $k_1$ and $k_2$). Neglecting the
part $\mu_a$ we find from (27) and (51)
\beq
R(Y,y_1,y_2)=\frac{2}{15\ln 2}
\frac{\langle r^2\rangle_P}{R_P^2}
\sqrt{\frac{\beta}{\pi}\frac{(Y-y_1)(Y-y_2)}{2Y-y_1-y_2}}
\eeq
If the two jets, forward and backward in the c.m. system, are taken
symmetric, that is,
\beq
y_1=\frac{1}{2}Y+y,\ \ y_2=\frac{1}{2}Y-y
\eeq
Eq. (53) becomes
\beq
R(Y,y)=\frac{1}{15\ln 2}
\frac{\langle r^2\rangle_P}{R_P^2}
\sqrt{\frac{\beta}{\pi}\left(Y-\frac{4 y^2}{Y}\right)}
\eeq

With the rapidity gap between the jet fixed and growing energy
the term $4y^2/Y$ can be neglected, so that the right-hand side
of (55) becomes independent of the gap $2y$ and proportional to $\sqrt{Y}$.
This means that eventually it inevitably becomes greater than unity,
so that the correlations become positive. The concrete threshold for
this depends on the values of $\bar{\alpha}$ and ratio
$\langle r^2\rangle/R_P^2$. With the Yukawa distribution of colour in the
projectile the last ratio is equal 2. Taking $\alpha_s=0.2$ we get for
(55)
\beq
R\simeq 0.195\sqrt{Y}
\eeq
So with these values correlations remain negative up to  quite high
energies corresponding to $Y<25$.

Note that in the framework of the old Regge-Gribov theory, due to the
assumed relations between the parameters and smallness of the
intercept $\Delta$, it was argued that the ratio $R$ for the part
$\mu_b$ is identically equal to unity, so that correlations are
entirely generated by the part $\mu_a$ [9].
In the present approach this does not hold. Although the factors
exponential in rapidities are indeed the same in $\mu_b$ and the
product $\mu(y_1,k_1)\mu(y_2,k_2)$, the power factors are not, so
that their final $Y$-dependence is different. Also the relation
between the coefficients is not fixed but depends on $\alpha_s$.
As a result the ratio $R$ for $\mu_b$ is different from unity and
depends on energy and $\alpha_s$, so that the correlations become
mainly determined by $\mu_b$ and thus much stronger than assumed
in earlier studies.

\section{Conclusions}
We have studied the inclusive jet production off the nucleus in the
perturbative QCD with $N_c\rightarrow$ infinity, in which the
total hA amplitude is exactly given by a sum of fan diagrams
constructed of BFKL pomerons. We have used the numerical results
obtained for this sum in our earlier paper [6].
Our main results are the following.

Our formulas confirm interpretation of function $h(y,q,b)$  defined by Eq.
(13) as the gluon density in the nucleus. In [6] this was done by comparison
with the perturbative expression for the structure function, the validity
of which for large values of the gluon density can be questioned.

The inclusive jet production rate is found to rapidly grow  with energy
in the same manner as the cross-section generated by a single pomeron
exchange. This result is an immediate consequence of the saturation of the
total hA cross-section, and was also found in earlier studies of the
fan diagrams with local pomerons.

However in contrast to these older studies, our jet production rate
is $A$ dependent: jet multiplicities grow roughly as $A^{2/9}$.
So the $A$ dependence is somewhat reduced as compared to the
eikonal result $A^{1/3}$, but not so strongly as thought before.
This result may have direct consequences for particle production
in nucleus-nucleus collisions, where we may expect the multiplicities
grow as $A^{10/9}$ for identical nuclei,
less than in the eikonal approach ($A^{4/3}$) but greater than in the
older fan diagram estimates ($A^{2/3}$).

The double inclusive jet production is dominated by the contribution
from Fig. 2b. Unlike older studies, this contribution is not
cancelled in the correlation function and leads to correlations which
start negative at lower energies and then change to
positive at a certain (rather high) energy. Their magnitude 
corresponds to a double pomeron exchange and so
grows very large at high energies.

In conclusion we want to remark that since the gluon distribution in
the nucleus is strongly shifted to higher momenta, the notorious
difficulty of the diffusion towards very small momenta seems to be
overcome. This problem
always raised doubts of the validity  of the perturbative treatment
of hadronic amplitudes: even if the initial wave function is centered
at high-momenta, its evolution in rapidity 
inevitably introduces low momentum contributions, for which the perturbative
treatment is not valid. The large nucleus seems to damp these small
momenta contributions, so that the use of the pertubation theory becomes
justified.

\section{References}

1. A.Mueller, Nucl. Phys.,{\bf B415} (1994) 373.\\
2. A.Mueller and B.Patel, Nucl. Phys.,{\bf B425} (1994) 471.\\
3. M.A.Braun and G.P.Vacca, Eur. Phys. J {\bf C6} (1999) 147.\\
4. Yu. Kovchegov, Phys. Rev {\bf D60} (1999) 034008.\\
5. I.Balitsky, hep-ph/9706411; Nucl. Phys. {\bf B463} (1996) 99.\\
6. M.A.Braun, hep-ph/0001268\\
7. Yu. Kovchegov, preprint CERN-TH/99-166 (hep-ph/9905214).\\
8. E.Levin and K.Tuchin, preprint DESY 99-108, TAUP 2592-99
(hep-ph/9908317).\\
9. A.Schwimmer, Nucl. Phys. B94 (1975)445.\\
10. L.N.Lipatov in: "Perturbative QCD", Ed. A.H.Mueller, World Sci.,
Singapore (1989) 411.\\

\section{Figure captions}
\noi Fig. 1.  The amplitude for hA scattering. Lines denote BFKL pomerons.\\
Fig. 2. Inclusive jet production from the lower (a) and upper (b) pomerons.\\
Fig. 3. Double inclusive jet production from the upper pomeron (a) and
immediately after the first spliting (b).\\
\newpage
\unitlength=1.00mm
\special{em:linewidth 0.4pt}
\linethickness{0.4pt}
\begin{picture}(159.00,142.00)
\put(14.67,140.33){\line(0,-1){23.67}}
\put(35.33,142.00){\line(0,-1){11.67}}
\put(35.33,130.33){\line(-1,-2){7.00}}
\put(35.33,130.33){\line(1,-2){7.33}}
\put(58.33,142.00){\line(0,-1){9.00}}
\put(68.00,133.33){\line(0,0){0.00}}
\put(58.33,133.33){\line(-1,-2){8.67}}
\put(54.67,125.67){\line(3,-5){5.67}}
\put(58.67,132.67){\line(2,-3){11.00}}
\put(114.67,121.67){\circle{14.00}}
\put(114.33,129.00){\line(0,1){12.67}}
\put(22.67,132.33){\line(0,-1){5.33}}
\put(20.67,129.67){\line(1,0){4.00}}
\put(42.67,129.67){\line(1,0){5.33}}
\put(45.33,132.00){\line(0,-1){4.67}}
\put(73.33,129.67){\line(1,0){4.33}}
\put(77.67,129.67){\line(1,0){0.33}}
\put(76.00,132.33){\line(0,-1){5.33}}
\put(81.67,124.00){\circle*{1.49}}
\put(85.00,124.00){\circle*{1.49}}
\put(88.00,124.00){\circle*{1.49}}
\put(94.00,131.00){\line(1,0){6.00}}
\put(94.33,129.00){\line(1,0){5.67}}
\put(114.67,122.00){\makebox(0,0)[cc]{$\Phi$}}
\put(60.33,82.33){\makebox(0,0)[cc]{Fig. 1}}
\end{picture}
\newpage
\unitlength=1.00mm
\special{em:linewidth 0.4pt}
\linethickness{0.4pt}
\begin{picture}(94.00,100.0)(0.00,22.33)
\put(68.67,122.33){\line(0,-1){27.00}}
\put(68.67,95.67){\line(-3,-5){22.67}}
\put(68.67,95.33){\line(2,-3){25.33}}
\put(56.00,74.33){\line(1,-2){8.33}}
\put(64.33,37.67){\makebox(0,0)[lc]{Fig. 2a}}
\put(56.33,84.00){\line(1,0){10.00}}
\put(54.33,86.67){\makebox(0,0)[rc]{$y,k$}}
\end{picture}

\unitlength=1.00mm
\special{em:linewidth 0.4pt}
\linethickness{0.4pt}
\begin{picture}(94.00,122.33)
\put(68.67,122.33){\line(0,-1){27.00}}
\put(68.67,95.67){\line(-3,-5){22.67}}
\put(68.67,95.33){\line(2,-3){25.33}}
\put(56.00,74.33){\line(1,-2){8.33}}
\put(64.33,37.67){\makebox(0,0)[lc]{Fig. 2b}}
\put(62.00,107.00){\line(1,0){13.67}}
\put(75.67,107.00){\line(0,0){0.00}}
\put(78.00,109.33){\makebox(0,0)[lc]{$y,k$}}
\end{picture}
\newpage

\unitlength=1.00mm
\special{em:linewidth 0.4pt}
\linethickness{0.4pt}
\begin{picture}(109.00,116.33)
\put(52.33,59.33){\circle{14.00}}
\put(102.00,60.00){\circle{14.00}}
\put(76.67,116.33){\line(0,-1){30.33}}
\put(76.67,85.67){\line(-1,-1){20.67}}
\put(56.00,65.00){\line(0,0){0.00}}
\put(76.67,85.33){\line(1,-1){20.00}}
\put(70.33,106.00){\line(1,0){12.67}}
\put(70.67,97.00){\line(1,0){12.00}}
\put(86.67,107.67){\makebox(0,0)[lc]{$y_1,k_1$}}
\put(86.67,94.67){\makebox(0,0)[lc]{$y_2,k_2$}}
\put(51.67,58.67){\makebox(0,0)[cc]{$\Phi$}}
\put(102.00,59.67){\makebox(0,0)[cc]{$\Phi$}}
\put(71.33,26.33){\makebox(0,0)[cc]{Fig. 3a}}
\end{picture}

\unitlength=1.00mm
\special{em:linewidth 0.4pt}
\linethickness{0.4pt}
\begin{picture}(96.33,109.67)
\put(51.67,56.67){\circle{14.00}}
\put(89.33,56.67){\circle{14.00}}
\put(69.67,109.67){\line(0,-1){20.00}}
\put(69.67,90.00){\line(-3,-5){16.00}}
\put(69.67,90.00){\line(3,-5){16.33}}
\put(75.67,70.00){\line(1,0){12.67}}
\put(55.67,79.67){\line(1,0){12.33}}
\put(55.00,82.67){\makebox(0,0)[rc]{$y_1,k_1$}}
\put(90.00,73.00){\makebox(0,0)[lc]{$y_2,k_2$}}
\put(50.00,56.67){\makebox(0,0)[cc]{$\Phi$}}
\put(88.67,57.33){\makebox(0,0)[cc]{$\Phi$}}
\put(64.33,20.67){\makebox(0,0)[cc]{Fig. 3b}}
\end{picture}
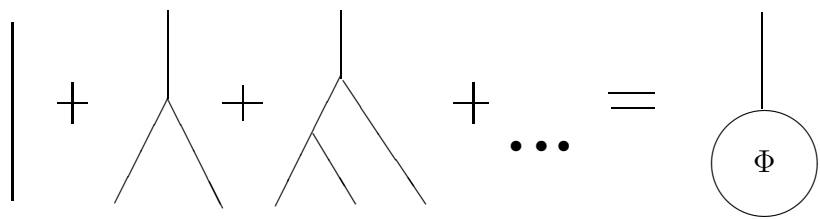
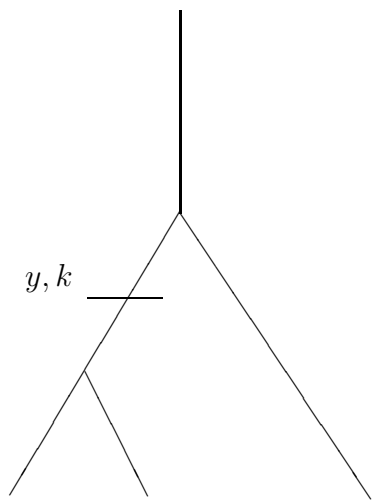
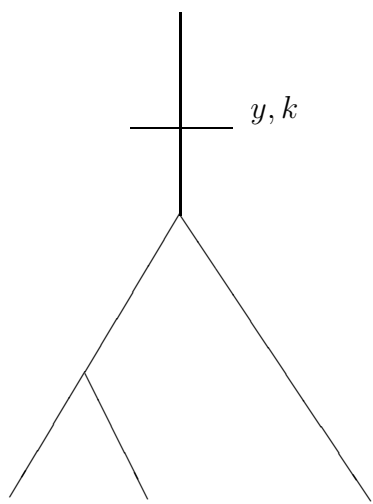
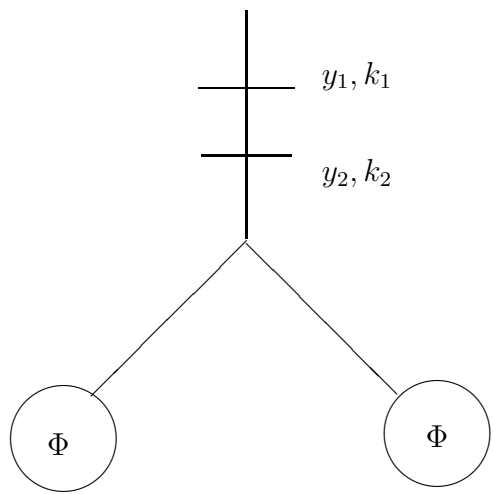
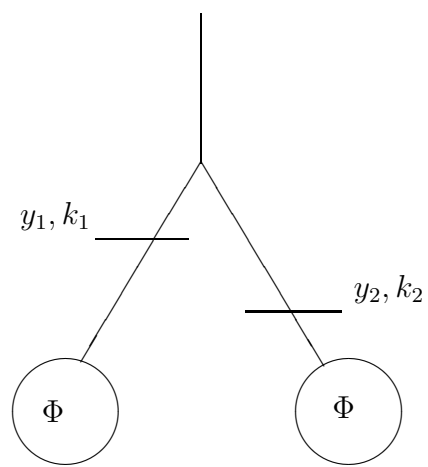
\end{document}